\documentclass[aps,prd,reprint,preprintnumbers,floats,epsfig,nofootinbib,amssymb,
nofootinbib]{revtex4}

\usepackage{slashed}
\usepackage{graphicx,color}
\usepackage{epsfig}
\usepackage{subfigure}
\usepackage{epsfig}
\usepackage{epstopdf}
\usepackage{dcolumn}
\usepackage{bm}
\usepackage{color}
\usepackage{ulem}
\usepackage{enumitem}
\usepackage{amsmath}
\definecolor{darkgreen}{rgb}{0.0,0.5,0.0}

\usepackage[linktocpage,
			colorlinks,
            linkcolor=black,
            anchorcolor=black,
            citecolor=black,
            urlcolor = darkgreen,
            ]{hyperref}

\begin{document}

\baselineskip=15pt


\title{Type-II Seesaw Triplet Scalar Effects
on Neutrino Trident Scattering}

\author{Yu Cheng$^1$, Xiao-Gang He$^2$, Zhong-Lv Huang$^3$, Ming-Wei Li$^1$}

\affiliation{\\}

\affiliation{$^1$
Tsung-Dao Lee Institute, and School of Physics and Astronomy\\ Shanghai Jiao Tong University, Shanghai 200240}

\affiliation{$^2$ 
National Center for Theoretical Sciences and Department of Physics\\ National Taiwan University, Taipei 10617}

\affiliation{$^3$ School of Physics, Xi’an Jiaotong University, Xi’an, 710049}

\begin{abstract}
In Type-II seesaw model,  an electroweak triplet scalar field $\Delta$ with a non-zero vacuum expectation value (vev) $v_\Delta$ is introduced to facilitate the generation of small neutrino masses. A non-zero $v_\Delta$ also affects the W mass through the electroweak $\rho$ parameter, making it to be less than 1 as predicted by standard model (SM). The component fields in $\Delta$ come along introduce additional contributions to reduce the SM rare neutrino trident scattering cross section. These fields also induce new processes not existed in SM,  such as $l_i \to \overline{ l_j} l_k l_l$ and $l_i \to l_j \gamma$. There are severe constraints on these processes which limit the effects on neutrino trident scattering and the $\rho$ parameter and therefore the W mass. The newly measured W mass by CDF makes the central value of $\rho$ parameter to be larger than 1, even larger than previously expected.  Combining neutrinoless double beta decay, direct neutrino mass and oscillation data, we find a lower limit for $v_\Delta$ as a function of the triplet scalar mass $m_\Delta$, $v_\Delta > (6.3 \sim 8.4) \mathrm{eV} (100 \mathrm{GeV}/m_\Delta)$. To have significant effect on $\rho$ in this model,  $v_\Delta$ needs to be in the range of a GeV or so. However this implies a very small  $m_\Delta$ which is ruled out by data. We conclude that the effect of triplet vev $v_\Delta$ on the W mass can be neglected. We also find that at 3$\sigma$ level, the deviation of the ratio for Type-II Seesaw to SM neutrino trident scattering cross section predictions is reduced to be below 1, but is restricted to be larger than 0.98. 
\end{abstract}

\maketitle

\noindent{\bf Introduction}

The establishment of neutrino oscillation implies that neutrinos have masses~\cite{ParticleDataGroup:2020ssz}. This is the first confirmed beyond standard model (SM) effect. To explain non-zero, yet small neutrino masses, new physics is needed. Type-II seesaw is among the most interesting ones~\cite{Lazarides:1980nt,Mohapatra:1980yp,Konetschny:1977bn,Cheng:1980qt,Magg:1980ut,Schechter:1980gr}. In Type-II seesaw model, an electroweak triplet scalar field $\Delta$, with its neutral component $\Delta^0$ has a non-zero vacuum expectation value (vev) $v_\Delta$ is introduced to facilitate the generation of small neutrino masses. 

A non-zero $v_\Delta$ will affect the W mass through the electroweak $\rho$ parameter  with $\rho = 1 - 2v^2_\Delta/(v^2+2 v^2_\Delta)$. Here $v=246$ GeV is the vev of the SM doublet Higgs field. While in the SM $\rho =1$. The recent new measurement~\cite{CDF:2022hxs} of $80,433.5\pm 9.4$ MeV for the W mass is 7$\sigma$ level above the SM prediction $80,357\pm6$ MeV. This makes the central value of $\rho$ to $1.0019$  while keep other SM parameters unchanged as before, which is larger than SM prediction and also larger than expected before~\cite{ParticleDataGroup:2020ssz}. This puts stringent constraint on the allowed $v_\Delta$. 

Because  $\Delta$ is a triplet transforming under the $SU(3)_c \times SU(2)_L\times U(1)_Y$ as $(1,\;3,\;1)$, its component field $\Delta^+$ introduces additional contributions to reduce the SM rare neutrino trident scattering process,  a muon-neutrino $\nu_\mu$ scatter off a heavy nuclei N to produce a pair of muon $\mu \bar \mu$, $\nu_\mu N \to N + \mu \bar \mu \nu_i$. Neutrino trident scattering is an interesting probe to new physics beyond SM. CHARM-II, CCFR and NuTeV  experiments have measured the cross section $\sigma$ for this process. The ratio, $\sigma/\sigma_{SM}$, the cross section $\sigma$ by experimental measurements to the SM predicted cross section $\sigma_{SM}$ are $1.58\pm 0.57$\cite{CHARM-II:1990dvf}, $0.82\pm 0.28$\cite{CCFR:1991lpl} and $0.72^{+1.73}_{-0.72}$\cite{NuTeV:1999wlw}, respectively. The average value is given by $\sigma_{\exp } / \sigma_{\mathrm{SM}}=0.95 \pm 0.25$. This is consistent with SM prediction within error bars although the central value is lower than the SM. Future improved data from experiments like DUNE\cite{Ballett:2018uuc} and $\nu$STORM\cite{Soler:2015ada} will provide more precise data to test SM prediction. Even the errors are relatively large at present, due to the rareness of the neutrino trident process, the ratio $\sigma/\sigma_{SM}$ is very sensitive to new physics, which has played an important role in constraining new physics beyond SM~\cite{Altmannshofer:2014pba,Altmannshofer:2016brv, Cen:2021iwv, Cheng:2021okr}. 

The component fields in $\Delta$ also induce new processes that not exist in the SM,  such as $l_i \to \overline{ l_j} l_k l_l$ and $l_i \to l_j \gamma$. There are severe constraints on these processes~\cite{ParticleDataGroup:2020ssz} which restrict the effects on neutrino trident scattering and the $\rho$ parameter and therefore the W mass. The model parameters are also constrained by neutrinoless double beta decay, direct neutrino mass and oscillation data. In this work we combine these data to study the effects of $\Delta$ on neutrino trident scattering and the constraint on vev $v_\Delta$. We find a lower limit for $v_\Delta$ as a function of the triplet scalar mass $m_\Delta$, $v_\Delta > (6.3 \sim 8.4) \mathrm{eV} (100 \mathrm{GeV}/m_\Delta)$, and that the effect of triplet vev $v_\Delta$ on the W mass can be neglected. We also find that at 3$\sigma$ level, the deviation of the ratio $\sigma/\sigma_{SM}$ of Type-II Seesaw and SM neutrino trident scattering cross section predictions is restricted to be less than 2\%. In the following we provide some details.
\\

\noindent{\bf Contributions to neutrino trident scattering}

 Writing $\Delta$ in its component form, we have
\begin{eqnarray}
\Delta = \left ( \begin{array}{cc}
\Delta^+/\sqrt{2}&\;\;\Delta^{++}\\
\Delta^0&\;\; - \Delta^+/ \sqrt{2}
\end{array}
\right )\;.
\end{eqnarray}
The Lagrangian term relevant to neutrino masses is the Yukawa interaction of the triplet and the left handed lepton doublet $L_i = (P_{L} \nu_i,\;P_{L}e_j)^T$ transforming as $(1,\;2,\-1/2)$
\begin{eqnarray}
\label{Yukawa}
L_{Yukawa} = Y_{ij} \overline{L^c}_i P_{L} L_j \Delta = \overline{\nu^c}_i Y_{ij}P_{L} \nu_j \Delta^0 - {1\over \sqrt{2}}(\overline{\nu^c}_i Y_{ij}P_{L} e_j \Delta^+ + \overline{e^c}_i Y_{ij}P_{L} \nu_j \Delta^+) - \overline{e^c}_i Y_{ij}P_{L} e_j \Delta^{++}\;.
\end{eqnarray}
$Y =(Y_{ij})$ is a symmetric matrix. The non-zero vev $v_\Delta/\sqrt{2}$ of $\Delta^0$ will generate neutrino masses. The mass matrix is given by $M_\nu = 2 Y \langle \Delta_0 \rangle =\sqrt{2} (Y_{ij} v_\Delta) = (m_{ij})$. $M_\nu$ can be diagonalized by a similarity transformation 
$U^TM_\nu U= \hat M_\nu$. $U$ is identified as the neutrino mixing matrix $U_{PMNS}$. The mass of singly- and doubly- charged scalars fields $\Delta^+$ and $\Delta^{++}$ are, to the leading order, equal which will be indicated as $m_{\Delta}$. 

Exchanging $\Delta^+$ at tree level using vertices from $L_{Yukawa}$, the following operator will be generated
\begin{eqnarray}
{m_{ij}m^*_{kl}\over  m^2_\Delta v^2_\Delta} \overline{\nu^c}_i \gamma^{\mu} P_{L} e_j \overline{ e_l} \gamma_\mu P_{L} \nu^c_k = {m_{ij}m^*_{kl}\over 2 m^2_\Delta v^2_\Delta} \overline{ \nu_k} \gamma^{\mu} P_{L} \nu_i \overline{ e_l} \gamma_\mu P_{L} e_j\;.
\end{eqnarray}
In the above equation we have used Fierz transformation and the fact $\overline{ l^c_i}\gamma_\mu l^c_j = - \overline{ l_j}\gamma^\mu l_i$. This effective interaction will contribute to the trident neutrino scattering and modify the measured $\sigma/\sigma_{SM}$.

If there is no corrections to the SM contribution, the value for $\sigma/\sigma_{SM}$ is equal to 1. The contributions from the $L_{Yukawa}$ will modify the SM prediction. Its contribution to $\nu_\mu N \to N \mu \bar \mu \nu_\mu$ has different sign as that from the SM one. In addition, there are also contributions to $\nu_\mu N \to N \mu \bar \mu( \nu_e + \nu_\tau)$. Since experimentally, the flavors of the final neutrinos are not identified, one should sum over all final states. We arrived at the final modification given by the following
\begin{equation}
\frac{\sigma}{\sigma_{SM}}=\frac{\left(1+4 s_{w}^{2}-\frac{2 m_{w}^{2}}{g^{2}} \frac{\left|m_{\mu \mu}\right|^{2}}{m_{\Delta}^{2} v_{\Delta}^{2}}\right)^{2}+\left(1-\frac{2 m_{w}^{2}}{g^{2}} \frac{\left|m_{\mu \mu}\right|^{2}}{m_{\Delta}^{2} v_{\Delta}^{2}}\right)^2+2\left(\frac{2 m_{w}^{2}\left|m_{\mu \mu}\right|^{2}}{g^{2} m_{\Delta}^{2} v_{\Delta}^{2}}\right)^2\left(\frac{\left|m_{e \mu}\right|^2+\left|m_{\tau \mu}\right|^2}{\left|m_{\mu \mu}\right|^{2}}\right)}{\left(1+4 s_{w}^{2}\right)^{2}+1}
\label{root}
\end{equation}

It is clear that in type-II seesaw model, there are new contributions to trident neutrino scattering. These contributions tend to reduce the $\sigma/\sigma_{SM}$ away from 1. As new contributions are expected to reduce SM value, it is in the right direction. If the current central value persists, the new contribution can help to explain data.  
\\

\noindent{ \bf Lepton flavor violation constraints on model parameters}

The  $L_{Yukawa}$ interaction will also induce other new processes which are not possible in the SM. For example, exchanging $\Delta^{++}$ at the tree level, will induce new flavor violating processes, such as $l_i \to \overline{ l_j} l_k l_l$. At the one loop level, by exchanging $\Delta^{++}$ and $\Delta^+$ in the loop,  $l_i \to l_j \gamma$ will be induced. Experimental searches for these processes have been carried out with null results which put stringent constraints on the model parameters. To study the implications of Type-II seesaw model for trident neutrino scattering, one should take into account data for these processes. We list them in Table \ref{lfv}.

Exchanging $\Delta^{++}$ at tree level, one obtains  the following operator
\begin{equation}
\label{Deltapp}
 \frac{m_{ij} m_{kl}^{*}}{2 v_{\Delta}^2 m^2_\Delta}\overline{e^c_i}P_{L}e_j \overline{e_l} P_R e^c_{k} = \frac{m_{ij}m^*_{kl}}{4 v^2_{\Delta} m^2_{\Delta}} \overline{e_l} \gamma_{\mu}P_{L} e_j \overline{e_k} \gamma^{\mu}P_{L} e_i\;.
\end{equation}
The branching ratio from the above is given by
\begin{equation}
Br \left(l_{i}^{-} \rightarrow l_{j}^{+} l_{k}^{-} l_{l}^{-}\right)=\frac{1}{2\left(1+\delta_{k l}\right)} \frac{m_{l_{i}}^{5} \tau_i}{192 \pi^{3}}\left|\frac{Y_{i j} Y^{*}_{k l}}{m_{\Delta^{++}}^{2}}\right|^{2}
=\frac{1}{8\left(1+\delta_{k l}\right)} \frac{m_{l_{i}}^{5} \tau_i}{192 \pi^{3}}\left|\frac{m_{i j} m_{k l}^*}{v^2_{\Delta} m_{\Delta}^{2}}\right|^{2}
\end{equation}
where $\tau_i$ is the lifetime of the i-th charged lepton, and the Kronecker delta $\delta_{k l}$ accounts for possible two identical final states.

The branching ratio for $l_{i}^{-} \rightarrow l_{j}^{-} \gamma $ generated at one-loop diagrams by exchanging $\Delta^{++}$ and $\Delta^{+}$ can be written as 
\begin{equation}
Br\left(l_{i}^{-} \rightarrow l_{j}^{-} \gamma\right)
=
\frac{m_{l_{i}}^{5} \alpha_{e m} \tau_i}{\left(192 \pi^{2}\right)^{2}}
\left|(Y^{\dagger} Y)_{ji}\right|^{2}\left(\frac{1}{m_{\Delta^{+}}^{2}}+\frac{8}{m_{\Delta^{++}}^{2}}\right)^{2}
=
\frac{m_{l_{i}}^{5} \alpha_{e m} \tau_i}
{\left(192 \pi^{2}\right)^{2}}\left(\frac{9 \left|(m^{\dagger} 
	m)_{ji}\right|}{2 v_{\Delta}^2 m_{\Delta}^{2}}\right)^{2}
\end{equation}
In our calculation we take $\alpha_{em} = \alpha_{em}(m_Z) = 1/128$ .

These lepton flavor violation processes are strongly constrained by the experiment data.  From the present data, we can obtain the upper bound for some of $|Y_{j i} Y^{*}_{k l}|$.
For example, for $\tau \rightarrow 3 \mu$, which $Br(\tau \rightarrow 3 \mu) < 2.1 \times 10^{-8}$\cite{ParticleDataGroup:2020ssz}, one can obtain the following constraint
\begin{equation}
\left|Y^{*}_{\mu \mu}Y_{\tau \mu}\right| = \frac{|m^*_{\mu \mu} m_{\tau \mu}|}{2 v^2_{\Delta}}<8.01 \times 10^{-5}\left(\frac{m_{\Delta}}{100 \mathrm{GeV}}\right)^{2}\;.
\end{equation}
The other limits obtained using data for $l_i \to \overline{ l_j} l_k l_l$ are listed in Table \ref{lfv}. The $\mu \to \bar e ee$ provides the strongest constraint.

Using experimental bounds on $l_{i}^{-} \rightarrow l_{j}^{-} \gamma$ process, we can obtain additional limits for $\left| (Y^{\dagger} Y)_{ij}\right|$. We also list these limits in Table \ref{lfv}.  The most stringent limit is from $\mu \to e\gamma$.
\\

\begin{table}
	\begin{center}
		\begin{tabular}{lcl}
			\hline\hline
			Process & Branching ratio bound &  \multicolumn{1}{c}{Constraint}\\
			\hline 
			$\mu^- \to e^+e^-e^-$ & $1.0\times 10^{-12}$ & $\left|Y^{*}_{e e} Y_{\mu e}\right| 
			< 2.36 \times 10^{-7}\left(\frac{m_{\Delta}}{100 \mathrm{GeV}}\right)^{2}$\\
			$\tau^- \to e^+e^-e^-$ & $2.7\times 10^{-8}$ & $\left|Y^{*}_{e e}Y_{\tau e}\right| 
			< 9.08 \times 10^{-5}\left(\frac{m_{\Delta}}{100 \mathrm{GeV}}\right)^{2}$ \\
			$\tau^- \to e^+e^-\mu^-$ & $1.8\times 10^{-8}$ & $\left|Y^{*}_{e \mu}Y_{\tau e}\right| 
			< 5.24 \times 10^{-5}\left(\frac{m_{\Delta}}{100 \mathrm{GeV}}\right)^{2}$\\
			$\tau^- \to e^+\mu^-\mu^-$ & $1.7\times10^{-8}$ & $\left|Y^{*}_{\mu \mu}Y_{\tau e}\right| 
			< 7.21 \times 10^{-5}\left(\frac{m_{\Delta}}{100 \mathrm{GeV}}\right)^{2}$ \\
			$\tau^- \to \mu^+e^-e^-$ & $1.5\times 10^{-8}$ &$\left|Y^{*}_{e e}Y_{\tau \mu}\right| 
			< 6.77 \times 10^{-5}\left(\frac{m_{\Delta}}{100 \mathrm{GeV}}\right)^{2}$ \\
			$\tau^- \to \mu^+\mu^-e^-$ & $2.7\times 10^{-8}$ & $\left|Y^{*}_{\mu e} Y_{\tau \mu}\right| 
			< 6.42 \times 10^{-5}\left(\frac{m_{\Delta}}{100 \mathrm{GeV}}\right)^{2}$ \\
			$\tau^- \to \mu^+\mu^-\mu^-$ & $2.1\times 10^{-8}$ & $\left|Y^{*}_{\mu \mu}Y_{\tau \mu}\right|
			< 8.01 \times 10^{-5}\left(\frac{m_{\Delta}}{100 \mathrm{GeV}}\right)^{2}$\\[1ex]
			$\mu\to e\gamma$ & $4.2\times 10^{-13}$ & $\begin{cases}
				\left|\left(Y^{\dagger} Y\right)_{e \mu}\right| <2.36 \times 10^{-6}\left(\frac{m_{\Delta}}{100 \mathrm{GeV}}\right)^{2}\\
				v_{\Delta} > 6.25 \mathrm{eV}\left(100 \mathrm{GeV}/m_{\Delta}\right)
			\end{cases}$ \\[3ex]
			$\tau\to e\gamma$ & $3.3\times 10^{-8}$ & $\begin{cases}
				\left|\left(Y^{\dagger} Y\right)_{e \tau}\right| 
				<1.55 \times 10^{-3}\left(\frac{m_{\Delta}}{100 \mathrm{GeV}}\right)^{2}\\
				v_{\Delta} > 0.24 \mathrm{eV}\left(100 \mathrm{GeV}/m_{\Delta}\right)
			\end{cases}$ \\[3ex]
			$\tau\to \mu\gamma$ & $4.4\times 10^{-8}$ & $\begin{cases}
				\left|\left(Y^{\dagger} Y\right)_{\mu \tau}\right| <1.79 \times 10^{-3}\left(\frac{m_{\Delta}}{100 \mathrm{GeV}}\right)^{2}\\
				v_{\Delta} > 0.56 \mathrm{eV}\left(100 \mathrm{GeV}/m_{\Delta}\right)
			\end{cases}$\\
			& & \\
			\hline\hline
		\end{tabular}
		\caption{Constraints for $|Y_{i j} Y^{*}_{k l}|$, $\left| Y^{\dagger} Y\right|_{j i}$ and $v_{\Delta}$ from different lepton flavor violation decay processes. The lower bound for $v_{\Delta}$ is obtained by numerically finding the lowest value of $\left|U_{j 2} U_{i 2}^{*} \Delta m_{21}^{2}+U_{j 3} U_{i 3}^{*} \Delta m_{31}^{2}\right|$ in the $3 \sigma$ range for the mixing parameters which shown in Table.~\ref{mixing} and then solving the inequality for $v_{\Delta}$.}
		\label{lfv}
	\end{center}
\end{table}

\noindent {\bf Constraints from neutrino masses and mixing on model parameters}

We now constrain for model parameters using information from neutrino masses and mixing in combination with rare decays discussed before. We find an interesting relation setting a lower bound on the vev $v_\Delta$ by combining neutrino mass and mixing matrix $U$ from neutrino oscillation in Table \ref{mixing} and $l_i \to l_j \gamma$. After using the unitarity property of $U$, we have 
\begin{equation}
\label{uppervdelta}
\left|\left(Y^{\dagger} Y\right)_{j i}\right|=\frac{1}{2 v_{\Delta}^{2}}\left|(m^\dagger m)_{ji}\right| = \frac{1}{2 v_{\Delta}^{2}}\left|U_{j 2} U_{i 2}^{*} \Delta m_{21}^{2}+U_{j 3} U_{i 3}^{*} \Delta m_{31}^{2}\right|\;.
\end{equation}
Note that the above equation only depends on the mass squared difference which are known from neutrino oscillation. The first term is proportional to $\Delta m_{21}^{2}$ which is much smaller than $\Delta m_{31}^{2}$, and its contribution is much smaller.   Numerically, we can get the lower bound for $v_{\Delta}$ by finding the lowest value of $ \left|U_{j 2} U_{i 2}^{*} \Delta m_{21}^{2}+U_{j 3} U_{i 3}^{*} \Delta m_{31}^{2}\right|$ for a given $m_\Delta$ using data from $l_i \to l_j \gamma$. The results for $v_\Delta$ from $l_i\to l_j \gamma$ are also listed in Table \ref{lfv}.
The strongest constraint is from $\mu^- \rightarrow e^- \gamma$. Using the limit from Table \ref{lfv} and neglecting the first term, we obtain 
\begin{equation}
v_{\Delta}>4.60 \times 10^{2}\left|s_{13} s_{23} \Delta m_{31}^{2}\right|^{\frac{1}{2}}\left(\frac{100 \mathrm{GeV}}{m_{\Delta}}\right) \cong (6.25-8.39) \mathrm{eV}\left(\frac{100 \mathrm{GeV}}{m_{\Delta}}\right)
\label{vdelta}
\end{equation}
The bounds for for NO and IO cases are approximately the same. In the future,  sensitivity of MEG II for $\mu \rightarrow e \gamma$ process will reach $6 \times 10^{-14}$\cite{Chiappini:2021ytk},  a stronger limit on  $v_{\Delta} > 39 \mathrm{eV} ({100 \mathrm{GeV}}/{m_{\Delta}})$ will be obtained.

The above implies that there is a minimal effects on $\rho$ since it is given by $1 - 2v^2_\Delta/(v^2+4 v^2_\Delta)$. 
Previous constraint on $\rho$ parameter is given by~\cite{ParticleDataGroup:2020ssz} $\rho = 1.00038\pm0.00020$.
The recent new measurement~\cite{CDF:2022hxs} of $m^{CDF}_W = 80,433.5\pm 9.4$ MeV for the W mass is 7$\sigma$ level above the SM prediction $m^{SM}_W = 80,357\pm 6$ MeV. It is also larger than the previous averaged value\cite{ParticleDataGroup:2020ssz} $m^{PDG}_W = 80,367\pm 6$ MeV. With the new value for $m_W$, the central value of $\rho$ would be enhanced by a factor $(m^{CDF}_W/m^{PDG}_W)^2$, making $\rho = 1.0019$ which is bigger than before value.  But $v_\Delta$ in type-II seesaw model reduces $\rho$ moving in the modification in the wrong direction. However, significant  effect on $\rho$ needs $v_\Delta$ to be of order a GeV or so. This would require $m_\Delta$ too small to be compatible with experimental lower limit of a few hundred GeV~\cite{ParticleDataGroup:2020ssz}.  We therefore can conclude that the Type-II seesaw vev $v_{\Delta}$ contribution to $m_W$ is negligible.  But Type-II seesaw model can have other effects at loop level such as exchange $\Delta$ component fields which mix with the SM Higgs boson to affect $m_W$ which should be studied further.  

We comment that if in the model there are multi-scalars $H_i$ with vevs $v_i$, one would have $\rho = \sum_i (I_i(I_i+1) - Y^2_i)v^2_i/2 \sum_i Y^2_i v^2_i$, where $I_i$ and $Y_i$ are the weak isospin and hypercharge of $H_i$. If the reason for a $\rho$ larger than 1 is due to new scalars, their hypercharge to be zero will maximize the effects. An example is to add a new triplet $\xi$ transforming as $(1,\;3,\;0)$ in our model. In this case $\rho = 1 - 2v^2_\Delta/(v^2+4 v^2_\Delta)  +4v^2_\xi/(v^2+4 v^2_\Delta)$~\cite{Cen:2018wye}. To produce a deviation of 0.0019 for $\rho -1$,  $v_\xi$ needs to be about  5.4 GeV.

\begin{table}
	\begin{center}
		\begin{tabular}{|l|cc|cc|}
			\hline & \multicolumn{2}{c|}{ Normal Ordering } & \multicolumn{2}{c|}{ Inverted Ordering} \\
			\cline { 2 - 5 } & $\mathrm{bfp} \pm 1 \sigma$ & $3 \sigma$ range & $\mathrm{bfp} \pm 1 \sigma$ & $3 \sigma$ range \\
			\hline $\sin ^{2} \theta_{12}$ & $0.304_{-0.012}^{+0.013}$ & $0.269 \rightarrow 0.343$ & $0.304_{-0.012}^{+0.012}$ & $0.269 \rightarrow 0.343$ \\
			$\theta_{12} /{ }^{\circ}$ & $33.44_{-0.74}^{+0.77}$ & $31.27 \rightarrow 35.86$ & $33.45_{-0.74}^{+0.77}$ & $31.27 \rightarrow 35.87$ \\
			$\sin ^{2} \theta_{23}$ & $0.573_{-0.023}^{+0.018}$ & $0.405 \rightarrow 0.620$ & $0.578_{-0.021}^{+0.017}$ & $0.410 \rightarrow 0.623$ \\
			$\theta_{23} /{ }^{\circ}$ & $49.2_{-1.3}^{+1.0}$ & $39.5 \rightarrow 52.0$ & $49.5_{-1.2}^{+1.0}$ & $39.8 \rightarrow 52.1$ \\
			$\sin ^{2} \theta_{13}$ & $0.02220_{-0.00062}^{+0.0008}$ & $0.02034 \rightarrow 0.02430$ & $0.02238_{-0.00062}^{+0.00064}$ & $0.02053 \rightarrow 0.02434$ \\
			$\theta_{13} /{ }^{\circ}$ & $8.57_{-0.12}^{+0.13}$ & $8.20 \rightarrow 8.97$ & $8.60_{-0.12}^{+0.12}$ & $8.24 \rightarrow 8.98$ \\
			$\delta_{\mathrm{CP} /}^{\circ}$ & $194_{-25}^{+52}$ & $105 \rightarrow 405$ & $287_{-32}^{+27}$ & $192 \rightarrow 361$ \\
			$\frac{\Delta m_{21}^{2}}{10^{-5} \mathrm{eV}^{2}}$ & $7.42_{-0.20}^{+0.21}$ & $6.82 \rightarrow 8.04$ & $7.42_{-0.20}^{+0.21}$ & $6.82 \rightarrow 8.04$ \\
			$\frac{\Delta m_{3 l}^{2}}{10^{-3} eV^2}$ & $+2.515_{-0.028}^{+0.028}$ & $+2.431 \rightarrow+2.599$ & $-2.498_{-0.029}^{+0.028}$ & $-2.584 \rightarrow-2.413$ \\
			\hline
		\end{tabular}
		\caption{Neutrino mixing parameters for best fit point and 3$\sigma$ range from the lastest NuFIT result\cite{Esteban:2020cvm}. Note that $\Delta m_{3 l}^{2}=\Delta m_{31}>0$ for $N O$ and $\Delta m_{3 l}=\Delta m_{32}<0$ for IO.}
		\label{mixing}
	\end{center}
\end{table}

There are also constraints for the $Y_{ij} = m_{ij}/\sqrt{2} v_\Delta$ from neutrino mass measurements. 
Due to the Majorana feature of neutrinos in type-II seesaw model, the neutrinoless double beta decay process can happen and current experiment result can constrain the model. Non-observation of neutrinoless double beta decay will constrain the allowed range for $m_{ee}$.
The strongest constraint is from KamLAND-Zen experiment\cite{KamLAND-Zen:2016pfg},which uses 13 Tons of Xe-loaded liquid scintillator to search for the neutrinoless double beta decay of $^{136}$Xe and has set a bound on the half-life of $T_{1 / 2}^{0 \nu}>1.07 \times 10^{26}$ yr at 90\% CL. The corresponding upper bound on the effective Majorana mass is  $m_{e e}<61-165 \mathrm{meV}$.
Here the region for upper bound of $m_{e e}$ is due to the theoretical uncertainties in the nuclear matrix element calculations. Also the limit on the lightest neutrino mass $m_{light}$ is obtained as (180-480) meV at 90\% C.L.

Expressing $m_{ee}$ in terms of the neutrino masses and mixing parameters, we have
\begin{eqnarray}
m_{ee} = \sqrt{2}Y_{ee} v_\Delta = m_1 U^2_{e1} + m_2 U^2_{e2} + m_3U^2_{e3}\;, 
\end{eqnarray}

\begin{figure}[!t]
	\centering
	\subfigure[\label{mee}]
	{\includegraphics[width=.486\textwidth]{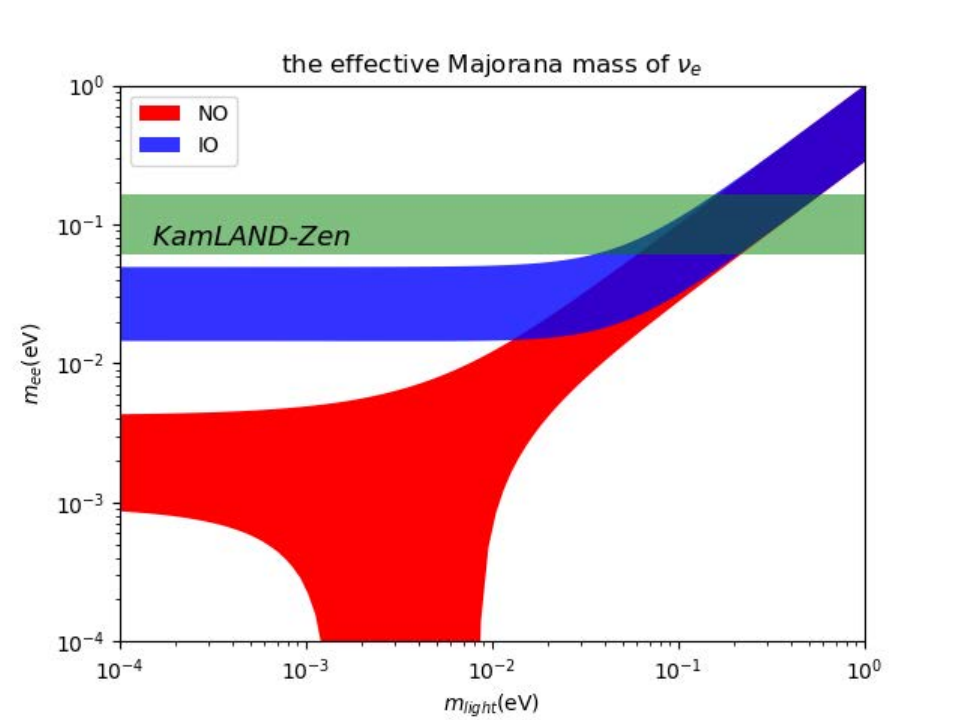}}
	\caption{ The allowed range for $m_{e e}$ region for NO and IO in 3$\sigma$ region of mixing parameters, the green dashed region is the upper limit range on the effective Majorana mass from KamLAND-Zen experiment.}
	\label{Trident}
\end{figure}

Note that $m_{ee}$ is related to the absolute neutrino mass. We use the lightest neutrino mass $m_{light}$ (for NO, $m_{light} = m_1$ and for IO, $m_{light} = m_3$) and the mass squared differences $\Delta m^2_{21}$ and $\Delta m^2_{31}$ as input parameters.
We show the allowed range in Fig.~\ref{Trident} for $m_{e e}$ value as a function of $m_{light}$  in the 3$\sigma$ range of mixing parameters space. We also show the constraint on the parameter space from KamLAND-Zen experiment data.  The sensitivity will approach $6 \times 10^{26}$ yr for  KamLAND2-Zen experiment\cite{Shirai:2017xff,Gando:2020cxo} which will put more stringent constraints on the parameter space. 

Other terrestrial experiments also give strong constraint on the absolute neutrino mass scale.  The best current bound is provided by the KATRIN experiment\cite{KATRIN:2021uub} with an upper limit of $m_{\nu} < 0.8$eV where $m_{\nu}^{2} = 2(Y^\dagger Y)_{ee} v^2_\Delta = \sum_{i}\left|U_{\mathrm{e} i}\right|^{2} m_{i}^{2}$. This constrains the lightest neutrino mass to be less that about 0.79 eV which is much weaker than the neutrinoless double beta decay. Cosmological observations provide a stringent limit on the sum of neutrino masses. The constraints provided by the Planck collaboration obtain an upper limit of  $\sum m_{i} <0.12 \mathrm{eV}$ at 95\% CL\cite{Planck:2018vyg}. 
The best one is from Ref.\cite{Palanque-Delabrouille:2019iyz} which combined Lyman-$\alpha$ and Planck data with $\sum m_{i} <0.09 \mathrm{eV}$ at 95\% CL.
\\

\noindent {\bf Numerical analysis neutrino trident scattering}

\begin{figure}[!t]
	\centering
	\subfigure[\label{Trident_lower_Mu3e}]
	{\includegraphics[width=.486\textwidth]{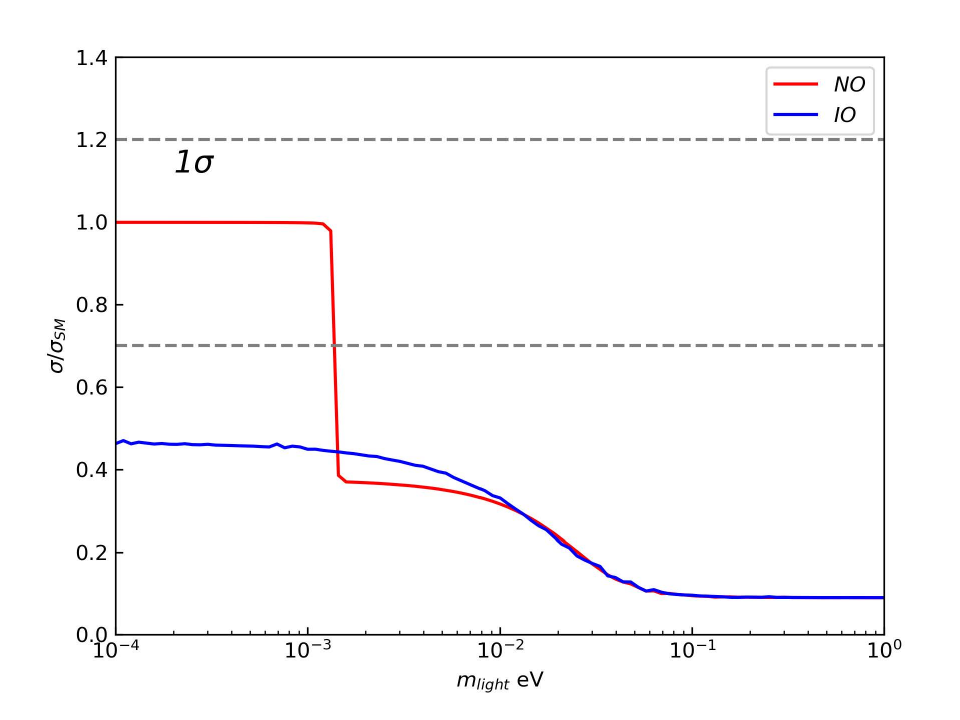}}
	\subfigure[\label{Trident_lower}]
	{\includegraphics[width=.486\textwidth]{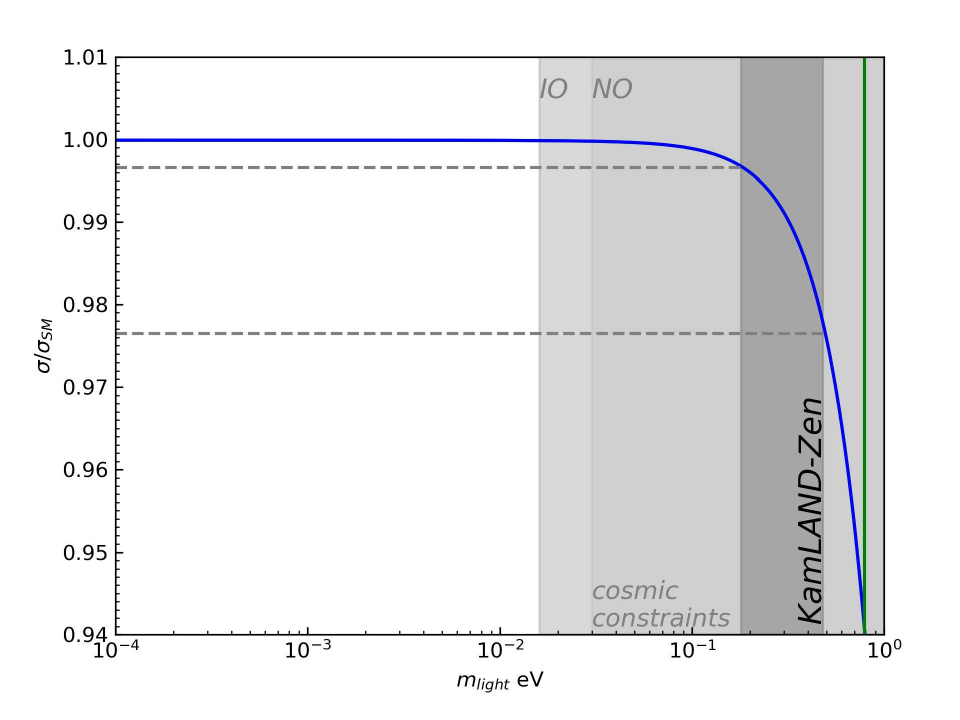}}
	\caption{ (a) The allowed range for $\sigma/\sigma_{SM}$ with $3\sigma$ constraints from $l_i \to \overline{ l_j} l_k l_l$. The region above the red(blue) line is allowed for NO(IO). The region between the two gray dashed lines represents for the 1$\sigma$ allowed region for neutrino trident process.  (b) The blue line shows the maximum deviation of $\sigma/\sigma_{SM}$ with $3\sigma$ constraints from $\mu  \to e \gamma$. The dark gray shaded area is the upper limit for lightest neutrino mass from KamLAND-Zen neutrinoless double beta decay data within unicertainties, and the region on the right hand side is excluded.
	The dashed gray lines represent the maximum deviation of $\sigma/\sigma_{SM}$ by considering the stronger(weaker) limit from KamLAND-Zen result. The region on the right hand side of the green line is excluded by KATRIN experimental data.
	The light gray shaded region is excluded by the comsic constraint from Ref.\cite{Palanque-Delabrouille:2019iyz}.}
	\label{Trident1}
\end{figure}

\begin{figure}[!t]
	\centering
	\subfigure[\label{meemeuovervd_NO}]
	{\includegraphics[width=.486\textwidth]{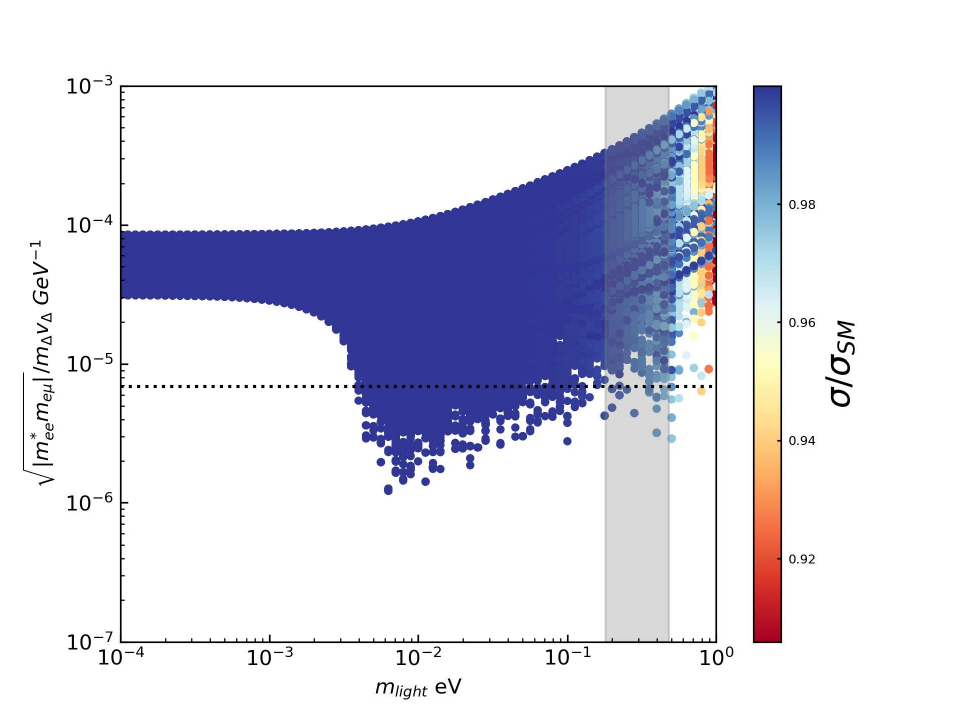}}
	\subfigure[\label{meemeuovervd_IO}]
	{\includegraphics[width=.486\textwidth]{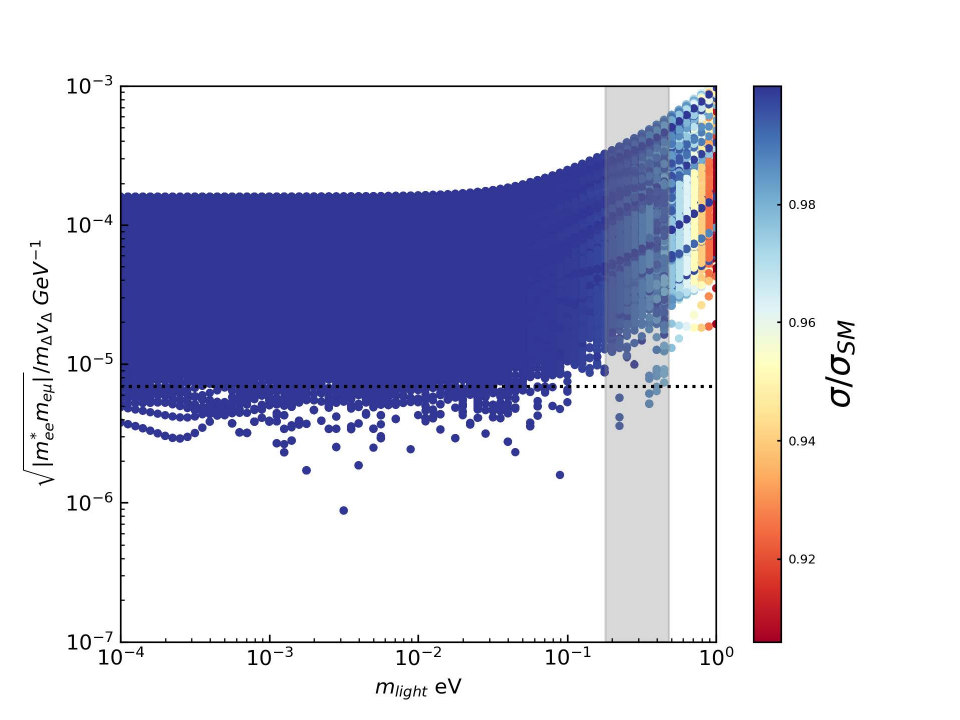}}
	\caption{$\sigma/\sigma_{S M}$ and $\sqrt{|m^*_{ee} m_{\mu e}|}/ m_{\Delta} v_{\Delta}$ value by scanning the parameter space of mixing angle and neutrino mass different in 3$\sigma$ region.The dashed line corresponds to the upper limit from $\mu \rightarrow \bar{e} e e$. Points above the dashed line are ruled out. The gray shaded area is the upper bound from KamLAND-ZEN experiment.
	(a) for NO case, and (b) for IO case.}
	\label{Global}
\end{figure}

We are now investigating the implications of Type-II seesaw model for  neutrino trident scattering from constraints discussed earlier. 
We will work with the allowed modifications to the trident neutrino scattering in the model taking the 3$\sigma$ ranges from various constraints. The results are shown in Figures \ref{Trident1} and \ref{Global}.

The processes $l_i \to \overline{ l_j} l_k l_l$ constrain some elements of $|Y_{ij} Y^*_{kl}|/m^2_\Delta$. Among them $\mu \to \bar e e e$ gives the strongest constraint. However, the elements $Y_{ij}$ come in different combinations for the contribution to neutrino trident scattering, there are room for large modifications to $\sigma/\sigma_{SM}$ as shown in (a) in Fig. \ref{Trident1}. Note here the $\sigma/\sigma_{SM}$ is constrained to be close to 1 for NO when $m_{light} < 10^{-3}$ while a steep decrease occur after that. This is because for NO case, the lower bound of  $\sqrt{|m^*_{ee} m_{\mu e }|}$ is constrained to be around $10^{-3}$ eV for $m_{light} < 10^{-3}$ and can reach a much smaller value when $m_{light}$ is bigger. So for $m_{light} > 10^{-3}$, we can choose a smaller value of $v_{\Delta}$ to make $|Y^*_{ee} Y_{\mu e}|/m^2_{\Delta} \equiv |m^*_{ee} m_{\mu e}|/ (2 v^2_{\Delta} m^2_{\Delta})$ satisfy the $\mu \to \bar e e e$ constraint and lead to a bigger modifications to $\sigma/\sigma_{SM}$ by using Eq.~\eqref{root}. Also we can see in Fig.~\ref{Trident_lower_Mu3e} that there are some regions for $\sigma/\sigma_{S M}$ to be larger than one from just this consideration, where the region between gray dashed lines represent for the 1$\sigma$ allowed region from neutrino trident experiment. These regions correspond to very small $v_{\Delta}$ and will be ruled out by $l_i \rightarrow l_j \gamma$ data as will be shown later. One may naively think that the constraints are not strong. This is true when consider $l_i \to \overline{ l_j} l_k l_l$ constraints alone. When combined with $l_i \to l_j \gamma$ which in general give more stringent constraints, it also rules out some parameter space allowed by $l_i \to l_j \gamma$ consideration. We discuss this now.

Among  $l_i \to l_j \gamma$, $\mu \to e \gamma$ gives the most stringent constraint. The blue soild line in Fig.~\ref{Trident_lower} shows the maximum deviation of $\sigma/\sigma_{SM}$ from 1 by considering only the $\mu \to e \gamma$ constraint from Eq.(\ref{uppervdelta}). We can find that $\sigma/\sigma_{SM}$ cannot be too far away for 1 for small $m_{light}$. When $m_{light}$ become of order an eV, the deviation from 1 can be significant. In this region, the neutrino masses are in the degenerate scenario. 

In the degenerate range of neutrino masses of order an eV, constraints from neutrinoless double beta decay from KamLAND-ZEN, shown as dark grey shaded region in Fig.~\ref{Trident_lower} which gives an upper limit on the lightest neutrino mass, and also tritium neutrino mass from KATRIN which exclude the region on the right of the green line in Fig.~\ref{Trident_lower}, become important. By taking into account these constraints, $\sigma/\sigma_{SM}$ can at most reach 0.98 (a few percent deviation from SM, shown as dashed gray line in the figure) at $3\sigma$ level. 
Future improved neutrinoless double beta decay KamLAND2-Zen will reach   $m_{ee} < 25-70$ meV which will put even more stringent constraint on $\sigma/\sigma_{SM}$.
Deviation up to a few percent level may be challenged by the sum of neutrino masses from cosmological considerations. We should keep this in mind. This limit on the sum from cosmological considerations on the other hand depend on cosmological models which need to be more carefully studied. If one takes the most stringent limit on the sum of neutrino mass constraint which is less than 0.09 eV, the effect of $\Delta$ on $\sigma/\sigma_{SM}$ is limited to be less than 0.1\%. It is a challenge to experimental test.


We mentioned that when combining $l_i\to \overline{ l_j} l_k l_l$ with $l_i\to l_j \gamma$ constraints, the former also plays important role as some of the parameter spaces satisfying $l_i\to l_j \gamma$ constraints cannot accommodate $\mu \to e \gamma$ constraint. We display the combined constraint for $\sigma/\sigma_{SM}$ in Fig. \ref{Global}. In the figure we vary $v_\Delta$ in the range $(6.3 \sim 20) \text{eV} (100 \text{GeV}/m_\Delta)$ to safely satisfy $l_i\to l_j \gamma$ constraints. We see that $\sigma/\sigma_{SM}$ is constrained to be less that 1 and the deviation from 1 is less than  about 2\% at $3\sigma$ level. Also $\mu \rightarrow \bar{e} e e$ rules out some region. 
\\

\noindent{ \bf Conclusions}

We have studied effects of Tyep-II seesaw triplet scalar and its vev on neutrino trident scattering and W mass through the vev modification to the electroweak parameter $\rho$. 
A non-zero $v_\Delta$ also affects the W mass through the electroweak $\rho$ parameter, making it to be less than SM prediction of 1. We find a lower limit for $v_\Delta$ as a function of the triplet scalar mass $m_\Delta$, $v_\Delta > (6.3 \sim 8.4) \text{eV} (100 \text{GeV}/m_\Delta)$. To have significant effect on $\rho$ in this model,  $v_\Delta$ needs to reach of order a GeV or so which implies a very small $m_\Delta$ not allowed by data. We conclude that the effect of triplet vev $v_\Delta$ on the W mass can be neglected.

The triplet scalar $\Delta$ in Type-II seesaw introduces additional contributions to reduce the SM rare neutrino trident scattering cross section. These fields also induce new processes not exist in the SM,  such as $l_i \to \overline{ l_j} l_k l_l$ and $l_i \to l_j \gamma$ which provide severe constraints to the model parameters. Combining these constraints with those coming from neutrinoless double beta decay, direct neutrino mass and oscillation data, we find that at 3$\sigma$ level the deviation of the ratio $\sigma/\sigma_{SM}$ is restricted to be larger than 0.98 which is closer to the current experimental central value compared with SM prediction. $\sigma/\sigma_{SM}$ will be probed by high sensitive experiments in the future. If a deviation is more than 2\% will be found, the Type-II seesaw is unlikely to be able to explain the data.

\section*{Acknowledgments}
This work was supported in part by Key Laboratory for Particle Physics, Astrophysics and Cosmology, Ministry of Education, and Shanghai Key Laboratory for Particle Physics and Cosmology (Grant No. 15DZ2272100), and in part by the NSFC (Grant Nos. 11735010, 11975149, and 12090064). XGH was supported in part by the MOST (Grant No. MOST 109-2112-M-002-017-MY3 ).

\providecommand{\href}[2]{#2}\begingroup\raggedright\endgroup


\end{document}